\documentclass[aps,twocolumn,amsmath,amssymb,floatfix,superscriptaddress]{revtex4-1}

\usepackage{graphicx}
\usepackage{color}
\usepackage{hyperref}

\begin{document}

\newcommand{\mrrm}[1]{\mathbf{P}[ #1 ]}

\title{Burst-tree decomposition of time series reveals the structure of temporal correlations} 
\author{Hang-Hyun Jo}
\email{hang-hyun.jo@apctp.org}
\affiliation{Asia Pacific Center for Theoretical Physics, Pohang 37673, Republic of Korea}
\affiliation{Department of Physics, Pohang University of Science and Technology, Pohang 37673, Republic of Korea}
\affiliation{Department of Computer Science, Aalto University, Espoo FI-00076, Finland}
\author{Takayuki Hiraoka}
\affiliation{Asia Pacific Center for Theoretical Physics, Pohang 37673, Republic of Korea}
\author{Mikko Kivel\"a}
\affiliation{Department of Computer Science, Aalto University, Espoo FI-00076, Finland}

\date{\today}

\begin{abstract}
    Comprehensive characterization of non-Poissonian, bursty temporal patterns observed in various natural and social processes is crucial to understand the underlying mechanisms behind such temporal patterns. Among them bursty event sequences have been studied mostly in terms of interevent times (IETs), while the higher-order correlation structure between IETs has gained very little attention due to the lack of a proper characterization method. In this paper we propose a method of decomposing an event sequence into a set of IETs and a burst tree, which exactly captures the structure of temporal correlations that is entirely missing in the analysis of IET distributions. We apply the burst-tree decomposition method to various datasets and analyze the structure of the revealed burst trees. In particular, we observe that event sequences show similar burst-tree structure, such as heavy-tailed burst size distributions, despite of very different IET distributions. The burst trees allow us to directly characterize the preferential and assortative mixing structure of bursts responsible for the higher-order temporal correlations. We also show how to use the decomposition method for the systematic investigation of such higher-order correlations captured by the burst trees in the framework of randomized reference models. Finally, we devise a simple kernel-based model for generating event sequences showing appropriate higher-order temporal correlations. Our method is a tool to make the otherwise overwhelming analysis of higher-order correlations in bursty time series tractable by turning it into the analysis of a tree structure.
\end{abstract}

\maketitle

\section{Introduction}\label{sec:intro}

A variety of dynamical processes in natural and social phenomena are known to be non-Poissonian or bursty, as observed in solar flares~\cite{Wheatland1998WaitingTime}, earthquakes~\cite{Corral2004LongTerm, deArcangelis2006Universality}, neuronal firings~\cite{Kemuriyama2010Powerlaw}, and human activities~\cite{Barabasi2005Origin, Karsai2018Bursty} to name a few. Traditionally, long-term temporal correlations have been characterized in terms of $1/f$ noise~\cite{Bak1987Selforganized, Weissman1988Noise, Ward2007Noise}, autocorrelation function~\cite{Kantelhardt2001Detecting, Allegrini2009Spontaneous, Karsai2012Universal, Yasseri2012Dynamics}, or Hurst exponent~\cite{Peng1994Mosaic,Kantelhardt2001Detecting, Allegrini2009Spontaneous, Rybski2009Scaling, Rybski2012Communication}. More recently, temporally correlated behavior, called bursts, has gained attention~\cite{Barabasi2005Origin, Karsai2018Bursty}. Bursts are rapidly occurring events in short-time periods alternating with long inactive periods. The mechanisms behind bursty temporal patterns have been studied by a number of modeling approaches~\cite{Barabasi2005Origin, Vazquez2006Modeling, Malmgren2008Poissonian, Malmgren2009Universality, Karsai2012Universal, Jo2013Contextual, Masuda2013Selfexciting, Wang2014Modeling, Jo2015Correlated, Garcia-Perez2015Regulation, Zipkin2016Pointprocess, Lee2018Hierarchical, Karsai2018Bursty}. It is also well-known that bursty interactions between elements of the systems influence the dynamical processes taking place in those systems, such as spreading or diffusion~\cite{Vazquez2007Impact, Karsai2011Small, Miritello2011Dynamical, Rocha2011Simulated, Jo2014Analytically, Delvenne2015Diffusion, Artime2017Dynamics, Hiraoka2018Correlated}. Therefore, characterization of the bursty temporal patterns is crucial to understand the underlying mechanisms for the emergent dynamics observed in various complex systems.

Temporal correlations in event sequences can be understood not only by statistical properties of time intervals between two consecutive events, i.e., interevent times (IETs), but also by correlations between IETs~\cite{Goh2008Burstiness, Jo2017Modeling}. The temporal correlations due to the heavy-tailed IET distributions have been extensively studied in recent years~\cite{Karsai2018Bursty}. In contrast, characterization and understanding of the sequence of IETs is far from being fully explored. The correlations between IETs have been described in terms of memory coefficient~\cite{Goh2008Burstiness} and bursty train size distribution~\cite{Karsai2012Universal} among others~\cite{Karsai2018Bursty}. The memory coefficient is a Pearson correlation coefficient between two consecutive IETs. To capture the structure beyond pairwise correlations the notion of bursty trains has been suggested. The size of bursty train or burst size is defined as the number of consecutive events that are not separated by IETs larger than some fixed time resolution or time window. Several empirical distributions of burst sizes are found to show heavy tails or power-law tails for a wide range of time windows~\cite{Karsai2012Universal, Wang2015Temporal}, which clearly implies the existence of higher-order correlations between IETs than simply expected by the memory coefficient~\footnote{Note that the exponential burst size distributions have recently been reported using the mobile phone communication dataset~\cite{Jiang2016Twostate}, implying the apparent absence of the higher-order correlations in the dataset analyzed.}. These findings naturally raise an important question: What is the origin of such higher-order temporal correlations? This issue has rarely been explored, except for a recent numerical study demonstrating the role of the tendency of bigger (smaller) bursts to be followed by bigger (smaller) ones in the higher-order temporal correlations~\cite{Jo2017Modeling}.

We stress that the burst size distribution of a time series is far from capturing the entire structure of temporal correlations present in the time series: The burst size distributions are often based on a few or even a single---often arbitrarily chosen---time resolutions, which limits the interpretation of the results to these specific resolutions. Further, information on the correlations between consecutive bursts is missing in the burst size distributions. This information is crucial for understanding the mechanisms behind the higher-order correlations between IETs evidenced by heavy-tailed burst size distributions, as well as for exploring the implications of this type of higher-order correlations. Therefore, it is strongly required to go beyond the current state of the art and devise a method for comprehensively characterizing the structure of temporal correlations in event sequences. 

In this paper we propose a method of decomposing any event sequence into an IET distribution and a burst tree. The IET distribution reveals the temporal scales between two consecutive events, while the burst tree does the same for their higher-order correlation structure. As the IET distributions have been extensively analyzed in the literature~\cite{Karsai2018Bursty}, we focus mostly on the higher-order correlation structure in the event sequences. By measuring various existing and newly introduced quantities from the revealed burst tree, such as the memory coefficient between consecutive bursts, we empirically demonstrate that the burst-tree decomposition is indeed useful to directly characterize the preferential and assortative mixing structure of bursts responsible for the higher-order correlations between IETs. Further, we observe that event sequences show similar burst-tree structure, such as heavy-tailed burst size distributions, despite of very different IET distributions. The burst-tree structure also allows us to construct novel microcanonical randomized reference models for event sequences~\cite{Gauvin2018Randomized}, which can be used to explore the implications of higher-order correlations in a controlled and meaningful way. Finally, we successfully generate event sequences showing the empirically observed higher-order temporal correlations using a simple model based on the burst-tree structure.

\section{Burst-tree decomposition method}\label{sec:method}

\begin{figure*}[!t]
    \includegraphics[width=\textwidth]{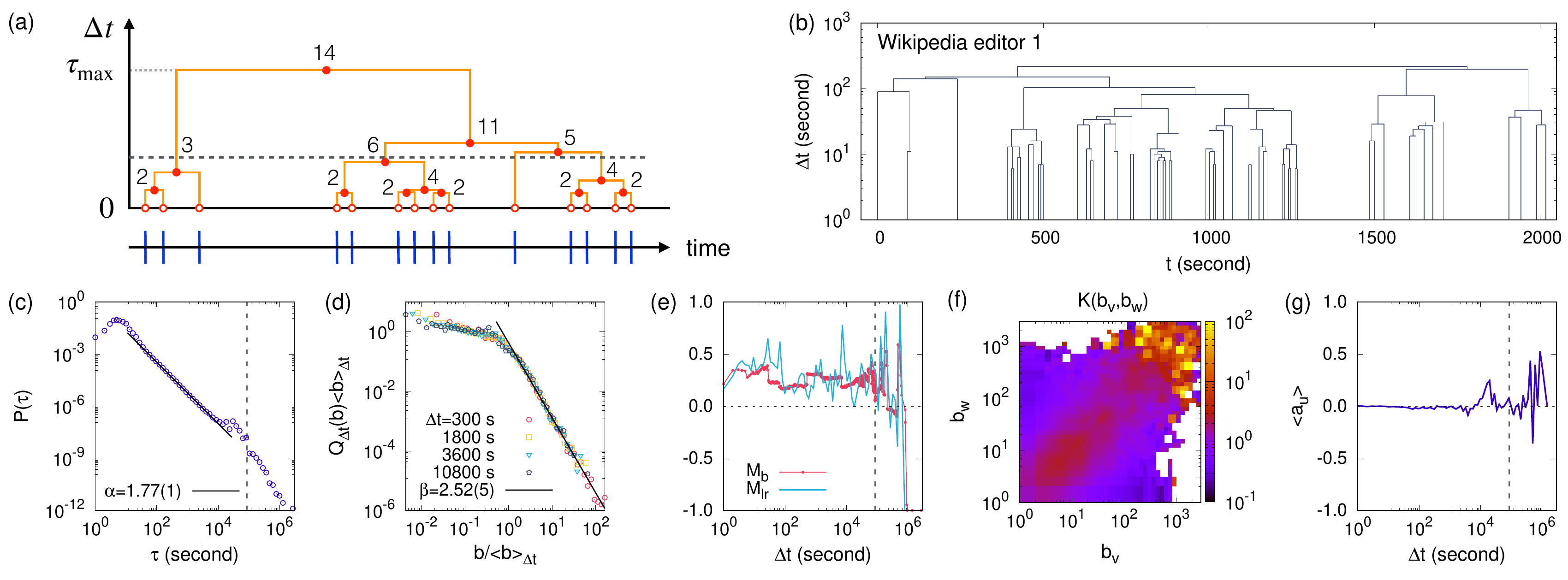}
    \caption{(a) Schematic diagram for the burst-tree decomposition method of an event sequence. The lower horizontal arrow denotes a time axis and blue vertical lines are events. The upper burst tree is derived by increasing the time window $\Delta t$ from $0$ to the maximum IET $\tau_{\rm max}$. At the bottom of the tree are the leaf nodes (red empty circles). Each internal node (red filled circle) in the tree denotes the merging of its left and right children, and the number next to the internal node is the burst size after merging. The height of the internal node corresponds to the IET between the last event of the left child and the first event of the right child. For the horizontal dashed line, see the main text. (b) A burst tree derived from a part of the edit sequence by the most active Wikipedia editor, namely, editor 1. (c--g) Empirical results of the editor 1's event sequence of $n\approx 1.1\times 10^6$ in terms of the IET distribution $P(\tau)$, burst size distributions $Q_{\Delta t}(b)$ for several values of $\Delta t$, memory coefficients between consecutive bursts $M_b$ and between sibling bursts $M_{lr}$ at each $\Delta t$, the merging kernel $K(b_v,b_w)$, and the averaged asymmetry $\langle a_u\rangle$ at each $\Delta t$, respectively. In the panel (d), $\langle b\rangle_{\Delta t}$ is the average burst size for a given $\Delta t$. Vertical dashed lines in panels (c,~e,~g) denote 1 day.}
    \label{fig:diagram_editor1}
\end{figure*}

We propose a burst-tree decomposition method for detecting the temporal correlation structure in an event sequence. For a given time window $\Delta t$, some consecutive events can be clustered to a bursty train: A bursty train is defined as a set of events such that interevent times (IETs) between any two consecutive events in the bursty train are less than or equal to $\Delta t$, while those between events in different bursty trains are larger than $\Delta t$~\cite{Karsai2012Universal}. The number of events in each bursty train is called a burst size. On the one hand, when $\Delta t$ is smaller than the minimum IET of the given event sequence, denoted by $\tau_{\rm min}$, each event constitutes a burst of size $1$ on its own. On the other hand, when $\Delta t$ is larger than the maximum IET, denoted by $\tau_{\rm max}$, all events belong to one burst, which we call a giant burst. Then by increasing $\Delta t$ continuously from $\tau_{\rm min}$ to $\tau_{\rm max}$, the bursts will consecutively merge to form bigger bursts, see Fig.~\ref{fig:diagram_editor1}(a) for a schematic diagram. Such a merging process can be fully described by a rooted tree whose leaf nodes, internal nodes, and the root node correspond to the events, the mergings or merged bursts, and the giant burst, respectively. Note that the root node is also an internal node. Hence, this burst tree of the event sequence reveals the hierarchical structure of temporal correlations~\footnote{The burst tree can be seen as a hierarchical data clustering in one-dimensional space using the nearest neighbor distance~\cite{Gan2007Data}.}.

We start by introducing a notation for event sequences. A given event sequence of $n+1$ events can be described by an ordered set of event timings $\{\hat t_0,\cdots,\hat t_n\}$. In most cases $\hat t_0$ indicates the beginning time of data collection. Otherwise, its relationship to the beginning time of data collection can be used to infer IET distributions~\cite{Kivela2015Estimating}. For convenience we shift timings by $\hat t_0$ using $t_i\equiv \hat t_i-\hat t_0$ for the $i$th event. This leads to the shifted event sequence $\mathcal{E}\equiv \{t_0,\cdots,t_n\}$ with $t_0=0$ by definition. From $\mathcal{E}$ the sequence of IETs is derived as $\{\tau_1,\cdots,\tau_n\}$ by the definition of $\tau_i\equiv t_i-t_{i-1}$. 

Using the shifted event sequence $\mathcal{E}$ we can now formally define the burst tree. Firstly, each of $n+1$ leaf nodes of the tree represents a single event, i.e., a burst of size $1$. Secondly, each of $n$ internal nodes of the tree, indexed by $u$, represents a merging of two consecutive bursts, indexed by $v$ and $w$, respectively. Here $v$ ($w$) is the index of the earlier (later) burst among them or the left (right) child of its parent node $u$. The IET between bursts $v$ and $w$, i.e., the time interval between the last event in $v$ and the first event in $w$, is associated with the internal node $u$, and this IET is denoted by $\hat\tau_u$. Note that the distribution of $\hat\tau_u$s is exactly the same as that of $\tau_i$s, denoted by $P(\tau)$. Although the leaf nodes are not associated with any IET by construction, we set their associated IETs as $0$ for convenience. The index $u$ for the internal node follows the rank of its associated IET in $\{\hat\tau_u\}$, e.g., $u=1$ for the root node as $\hat\tau_1=\tau_{\rm max}$. In sum, each internal node is represented by a tuple of $(u,v,w,\hat\tau_u)$, and the weighted burst tree by $\mathcal{T}\equiv \{(u,v,w,\hat\tau_u)\}$ for $u=1,\cdots,n$. Once the weighted burst tree is derived, the burst size for each internal node $u$, denoted by $b_u$, is computed as being equal to $b_v+b_w$. 

The weighted burst tree $\mathcal{T}$ is an alternative representation of the event sequence $\mathcal{E}$. That is, the event sequence $\mathcal{E}$ can be exactly reconstructed by visiting internal nodes of $\mathcal{T}$ in the inorder and by setting the $i$th event timing for $i=1,\cdots,n$ as $t_i=t_{i-1}+\hat\tau_{u(i)}$, where $u(i)$ denotes the $i$th visited internal node by the inorder traversal~\footnote{The inorder traversal is a depth-first way of traversing a tree by which one first visits the left (earlier) subtree before visiting the branching node and lastly the right (later) subtree.}. We denote this type of equivalence by $\mathcal{E} \widehat{=} \mathcal{T}$. Now $\mathcal{T}$ can be decomposed into the IET distribution $P(\tau)$ and the ordinal burst tree $\mathcal{G}\equiv \{(u,v,w)\}$. Here the ordinal burst tree retains the information on the ranks or orders of internal nodes, while the information on the IETs is discarded. As before, $\mathcal{T}$ can be exactly reconstructed by associating the $u$th largest IET in $P(\tau)$ to the internal node $u$ in $\mathcal{G}$. We denote this equivalence by $\mathcal{T} \widehat{=} (P(\tau),\mathcal{G})$. By transitivity, $\mathcal{E}$ is also equivalent to $(P(\tau),\mathcal{G})$, i.e., $\mathcal{E} \widehat{=} (P(\tau),\mathcal{G})$.

We note that if more than two consecutive bursts are separated by IETs of the same length, hence merged to the same node at the same time, then the order in which those bursts are merged in a pairwise manner is not well defined. In such cases, we randomly and uniformly choose two consecutive bursts and merge them into one burst, and repeat this binary merging until all these bursts are merged into one burst. These corner cases leading to the non-binary merging are insignificant for the analysis of the datasets in the next section~\footnote{See Supplemental Material at [URL] for the statistics of cases of merging more than two bursts at the same time for Wikipedia editor 1 (Sec.~I and Fig.~S1(a,~b)), empirical results of the 20 most active Wikipedia editors, the 20 most active Twitter users, and 20 healthy subjects (Sec.~II and Figs.~S2--S4), empirical results of percolation process for Wikipedia editor 1 (Sec.~III and Fig.~S1(c)), complete results of microcanonical randomized reference models for Wikipedia editor 1, Twitter user 1, heartbeat subject 1, and Japanese earthquakes (Sec.~IV and Figs.~S5--S8), and simulation results of other kernel-based models (Sec.~V and Figs.~S9 and~S10).}. We also remark that the burst-tree decomposition method has some conceptual resemblance to visibility graphs~\cite{Lacasa2008Time} in a sense that both methods map time series onto graphs.

\section{Higher-order structure in data}\label{sec:data}

\subsection{Data description}\label{subsec:data}

We consider four time series datasets: English Wikipedia, Twitter, heartbeat, and Japan University Network Earthquake Catalog (JUNEC). (i) We analyze edit sequences by editors from the English Wikipedia dump on October 2, 2015~\footnote{\url{https://dumps.wikimedia.org/}.}. Each edit is recorded with the timing of edit in a resolution of seconds. As a case study, we choose one of the most active editors, who edited more than $1.1$ million times for over $8.5$ years until 2015. We call this editor as the editor 1. (ii) We also analyze activity patterns of Twitter users in a dataset collected in 2009~\cite{Yang2011Patterns}. The data contains timings of tweets in a resolution of seconds. As an example, we focus on the most active user with around $87$ thousand tweets in the time period, which we call the user 1. (iii) We then analyze the heartbeat time series of healthy individuals or subjects (normal sinus rhythm) measured for 24-hour period in a resolution of milliseconds~\cite{Stein2003Normal}, which was downloaded from PhysioBank~\footnote{\url{https://physionet.org/physiobank/}.}. Here each event denotes each beat. We focus on one of the subjects, which we call the subject 1. (iv) Finally, we analyze the earthquake sequence in the JUNEC including around $2.0\times 10^5$ earthquakes occurred in Japan from July 1, 1985 to December 31, 1998~\footnote{\url{http://wwweic.eri.u-tokyo.ac.jp/db/junec/index.html}.}.

\subsection{Preferential and assortative mixing structure}\label{subsec:mixing}

As a case study, we analyze the Wikipedia editor 1's event sequence of $n\approx 1.1\times 10^6$ using the burst-tree decomposition method. In Fig.~\ref{fig:diagram_editor1}(b), we show a burst-tree structure derived from a part of the event sequence by the editor 1. We also find a power-law scaling in the interevent time (IET) distribution, i.e., $P(\tau)\sim \tau^{-\alpha}$ with $\alpha=1.77(1)$ for $30<\tau<10^4$ in seconds, as shown in Fig.~\ref{fig:diagram_editor1}(c). 

To study the mixing structure of bursts for the higher-order correlations between IETs, we measure various existing and newly introduced quantities. First of all, the burst size distribution $Q_{\Delta t}(b)$ for a time window $\Delta t$ simply reflects a horizontal cross-section of the weighted burst tree at the height of $\Delta t$, as depicted by crossing points between the edges of the tree and the horizontal dashed line in Fig.~\ref{fig:diagram_editor1}(a). Precisely, we collect pairs with a child (either leaf node or internal node) and its parent satisfying the condition that an IET associated with the child is smaller than or equal to $\Delta t$, while its parent is associated with an IET larger than $\Delta t$. Denoting the time-ordered set of such children by $C_{\Delta t}$, the burst size distribution $Q_{\Delta t}(b)$ is directly obtained from the burst sizes $b_u$ of nodes $u \in C_{\Delta t}$. We find for the editor 1 that $Q_{\Delta t}(b)\sim b^{-\beta}$ with $\beta=2.52(5)$ for a wide range of $\Delta t$, as shown in Fig.~\ref{fig:diagram_editor1}(d). To investigate the origin of such a power-law behavior in burst size distributions, we propose two definitions of memory coefficient between consecutive burst sizes, $M_b$ and $M_{lr}$, as well as the merging kernel $K(b_v,b_w)$. Later we also make use of the burst-tree structure to get some insight into the feature of time asymmetry by measuring asymmetries $\{a_u\}$.

The assortative mixing structure of bursts, i.e., a tendency of big (small) bursts to be followed by big (small) ones, can be directly tested by measuring the correlations between consecutive bursts. We define the memory coefficient $M_b$ for a given $\Delta t$ as a Pearson correlation coefficient between burst sizes of two consecutive nodes in $C_{\Delta t}$ as follows:
\begin{equation}
    \label{eq:memory_b}
    M_b \equiv \frac{1}{n_b - 1}\sum_{k=1}^{n_b-1}\frac{(b^{(k)} - \mu_1)(b^{(k+1)} - \mu_2)}{\sigma_1 \sigma_2},
\end{equation}
where $n_b = |C_{\Delta t}|$ is the number of bursts and $b^{(k)}$ denotes the burst size of the $k$th node in $C_{\Delta t}$ for $k=1,\cdots,n_b$. $\mu_1$ ($\mu_2$) and $\sigma_1$ ($\sigma_2$) denote the average and standard deviation of burst sizes of nodes except for the last (the first) node in $C_{\Delta t}$, respectively. Positive $M_b$ implies a tendency of big (small) bursts to be followed by big (small) ones. The opposite tendency can be observed for the negative $M_b$, while $M_b=0$ indicates the absence of correlations between consecutive burst sizes. Figure~\ref{fig:diagram_editor1}(e) shows that $M_b$ has positive values of $0.2\sim 0.3$ for several decades of $\Delta t$, clearly revealing the assortative mixing structure of bursts. Note that the value of $M_b$ fluctuates around $0$ for $\Delta t>2$ days, implying that the degrees of burstiness before and after the days without editing might be uncorrelated with each other.

\begin{figure*}[!t]
    \includegraphics[width=\textwidth]{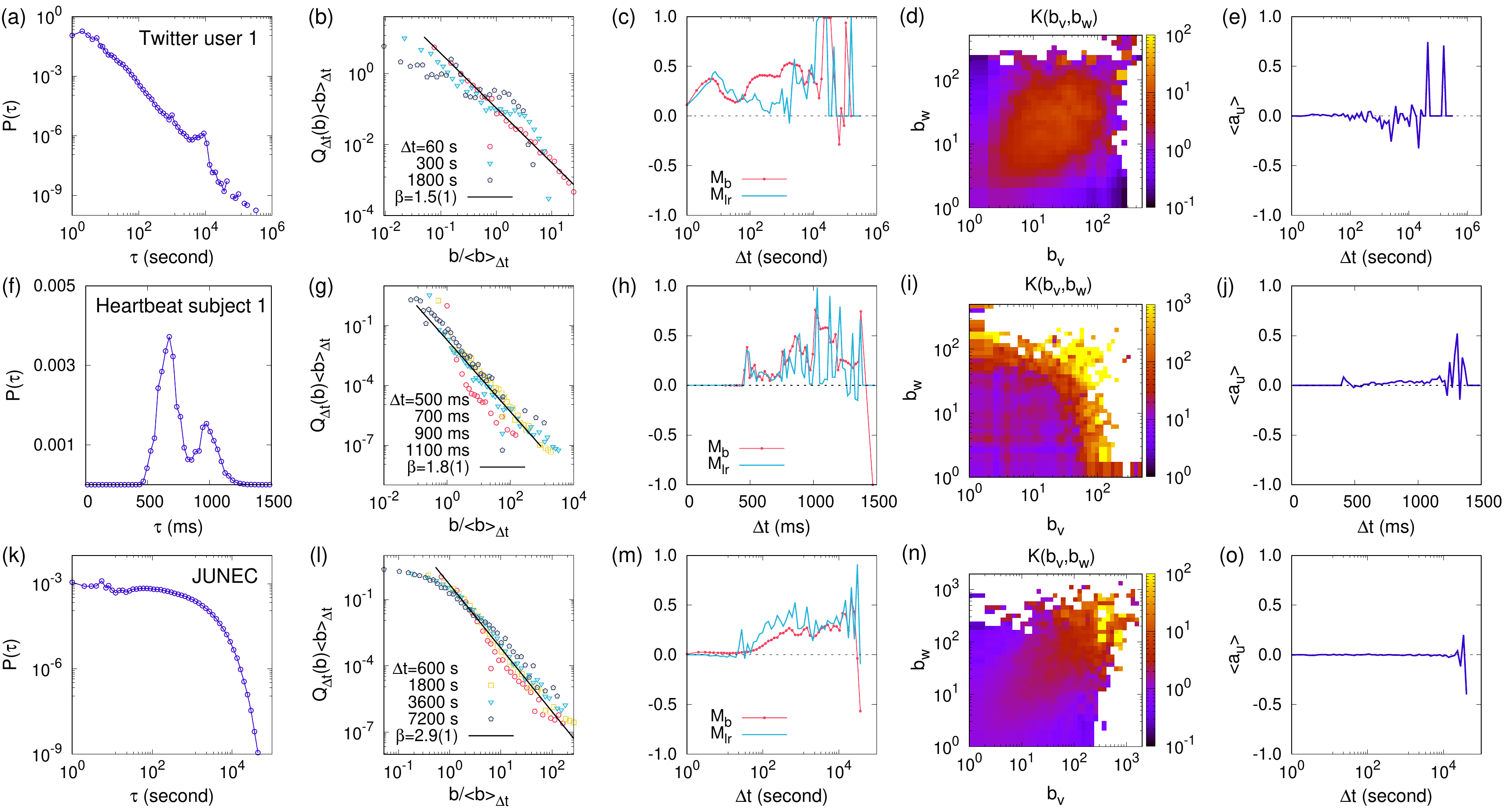}
    \caption{Empirical results for the tweet sequence of Twitter user 1 of $n\approx 8.5\times 10^4$ (top), the heartbeat time series of subject 1 of $n\approx 1.1\times 10^5$ (middle), and Japanese earthquake sequence (JUNEC) of $n\approx 2.0\times 10^5$ (bottom), in terms of the IET distribution $P(\tau)$, burst size distributions $Q_{\Delta t}(b)$, memory coefficients $M_b$ and $M_{lr}$, the merging kernel $K(b_v,b_w)$, and the averaged asymmetry $\langle a_u\rangle$ (from left to right), respectively. In panels (b,~g,~l), $\langle b\rangle_{\Delta t}$ is the average burst size for a given $\Delta t$.}
    \label{fig:other_data}
\end{figure*}

When measuring $M_b$ in Eq.~\eqref{eq:memory_b}, all pairs of two consecutive nodes or bursts in $C_{\Delta t}$ have been equally considered no matter how large IETs separating those bursts are, which may introduce some spurious correlations. Such spurious correlations can be corrected by considering only the pairs of sibling nodes in $C_{\Delta t}$, i.e., those sharing the same parent node. We denote the set of those siblings by $S_{\Delta t}\equiv \{(v,w)|v,w\in C_{\Delta t},\ (u,v,w) \in \mathcal{G}\}$. Then we suggest a novel definition of the memory coefficient between sibling bursts for a given $\Delta t$ as follows:
\begin{equation}
    \label{eq:memory_lr}
    M_{lr} \equiv \frac{1}{|S_{\Delta t}|}\sum_{(v,w)\in S_{\Delta t}}\frac{(b_v - \mu_l)(b_w - \mu_r)}{\sigma_l \sigma_r},
\end{equation}
where $\mu_l$ ($\mu_r$) and $\sigma_l$ ($\sigma_r$) respectively denote the average and standard deviation of burst sizes of all the left (right) children in $S_{\Delta t}$. In Fig.~\ref{fig:diagram_editor1}(e), we find that $M_{lr}$ has positive values of $0.1\sim 0.4$ for several decades of $\Delta t$, again showing the assortative mixing structure of bursts.

The correlations between siblings of burst sizes $b_v$ and $b_w$ for the whole burst tree can be more precisely characterized by estimating a merging kernel $K(b_v,b_w)$. This is based on the observation that the merging process with increasing $\Delta t$ from $\tau_{\rm min}$ to $\tau_{\rm max}$ can be interpreted as a stochastic process for coalescence in physical or networked systems~\cite{Aldous1999Deterministic}. Instead of $\Delta t$, we use the cumulative number of binary mergings, denoted by $s$, as an auxiliary time in the process. Then the merging process can be described as follows: At the time step $s=0$ (i.e., $\Delta t<\tau_{\rm min}$) we have $n+1$ events, equivalently, $n+1$ bursts of size $1$. At each time step $s$, one has $n+1-s$ bursts, whose burst size distribution is denoted by $Q_s(b)$. A pair of bursts among $n+1-s$ bursts are randomly chosen with a probability proportional to the merging kernel $K(b_v,b_w)$, and then they are merged into another burst of size $b_v+b_w$. This process is repeated until all events eventually belong to a giant burst (i.e., $\Delta t\geq \tau_{\rm max}$). 

Then we consider the empirical ordinal burst tree $\mathcal{G}$ as a realization of the above merging process. As each merging corresponds to an internal node $u$ in $\mathcal{G}$, the time step $s$ is related to the node index $u$ as $s=n-u$. For estimating the merging kernel from $\mathcal{G}$, we define $m_s(b_v,b_w)$ at the time step $s$, i.e., for the internal node $u=n-s$, as having a value of $1$ if its child nodes have burst sizes $b_v$ and $b_w$, and $0$ otherwise. The merging kernel is now estimated using the following formula:
\begin{equation}
    \label{eq:kernel_b}
    K(b_v,b_w) \equiv \frac{ \sum_{s=0}^{n-1} m_s(b_v,b_w)}{\sum_{s=0}^{n-1} Q_s(b_v) Q_s(b_w) }.
\end{equation}
Equation~\eqref{eq:kernel_b} has been modified from the formula that was introduced to numerically estimate the kernel for the preferential attachment in evolving scale-free networks~\cite{Pham2015PAFit}. Therefore, we expect the merging kernel to reveal the mechanism behind the power-law burst size distributions. 

From the merging kernel estimated for the editor 1 in Fig.~\ref{fig:diagram_editor1}(f) we make three important observations: (i) $K(b_v,b_w)$ shows a high profile in the diagonal part around the line of $b_v=b_w$, while it has low values in the off-diagonal part. (ii) The diagonal cross-section $K(b,b)$ is an overall increasing function of $b$. (iii) $K(b_v,b_w)$ shows an overall symmetric behavior with respect to the diagonal axis, implying $K(b,b')\approx K(b',b)$. The observation (i) is indeed consistent with positive $M_b$ and $M_{lr}$ in Fig.~\ref{fig:diagram_editor1}(e), confirming the assortative mixing of bursts. On the other hand, the observation (ii) implies the preferential mixing structure of bursts, by which the bigger bursts tend to be merged earlier, i.e., at smaller time scales, than the smaller ones. Conclusively, these empirical evidences enable us to understand the power-law behavior of burst size distributions in Fig.~\ref{fig:diagram_editor1}(d) by means of the preferential and assortative mixing structure. Further, a possible connection between the observation (iii) and the averaged asymmetry $\langle a_u\rangle$ in Fig.~\ref{fig:diagram_editor1}(f) will be discussed below. 

We demonstrate how the burst-tree structure can be used to study the time asymmetry in the event sequence regarding the issues of nonlinearity or irreversibility in time series~\cite{Theiler1992Testing, Daw2000Symbolic, Porporato2007Irreversibility, Donges2013Testing}. For this, we define for each internal node $u$ the asymmetry between its two children with burst sizes $b_v$ and $b_w$ as
\begin{equation}
    a_u \equiv \frac{b_w-b_v}{b_w+b_v}\,.
\end{equation}
If $b_w=b_v$, one gets $a_u=0$. If $b_w\gg b_v$, we have $a_u\approx 1$, while $a_u\approx -1$ for $b_w\ll b_v$. For a given $\Delta t$, we take an average of $a_u$ over the nodes whose associated IETs are the same as $\Delta t$, and denote it by $\langle a_u\rangle$. For the editor 1, the value of $\langle a_u\rangle$ turns out to remain close to $0$ for $\Delta t<2$ hours, while it becomes positive for $\Delta t$ between $2$ and $10$ hours, see Fig.~\ref{fig:diagram_editor1}(g). It implies that relatively small bursts tend to be followed by relatively big bursts after several hours of inactivity. One can interpret this observation such that the less bursty editing activities in the afternoon tend to be followed by the more bursty editing activities at night after several hours. Note that the behavior of $\langle a_u\rangle\approx 0$ for small $\Delta t$ can be to some extent considered as being consistent with the observation (iii) for the merging kernel, namely, $K(b,b')\approx K(b',b)$. 

\begin{figure*}[!t]
    \includegraphics[width=0.8\textwidth]{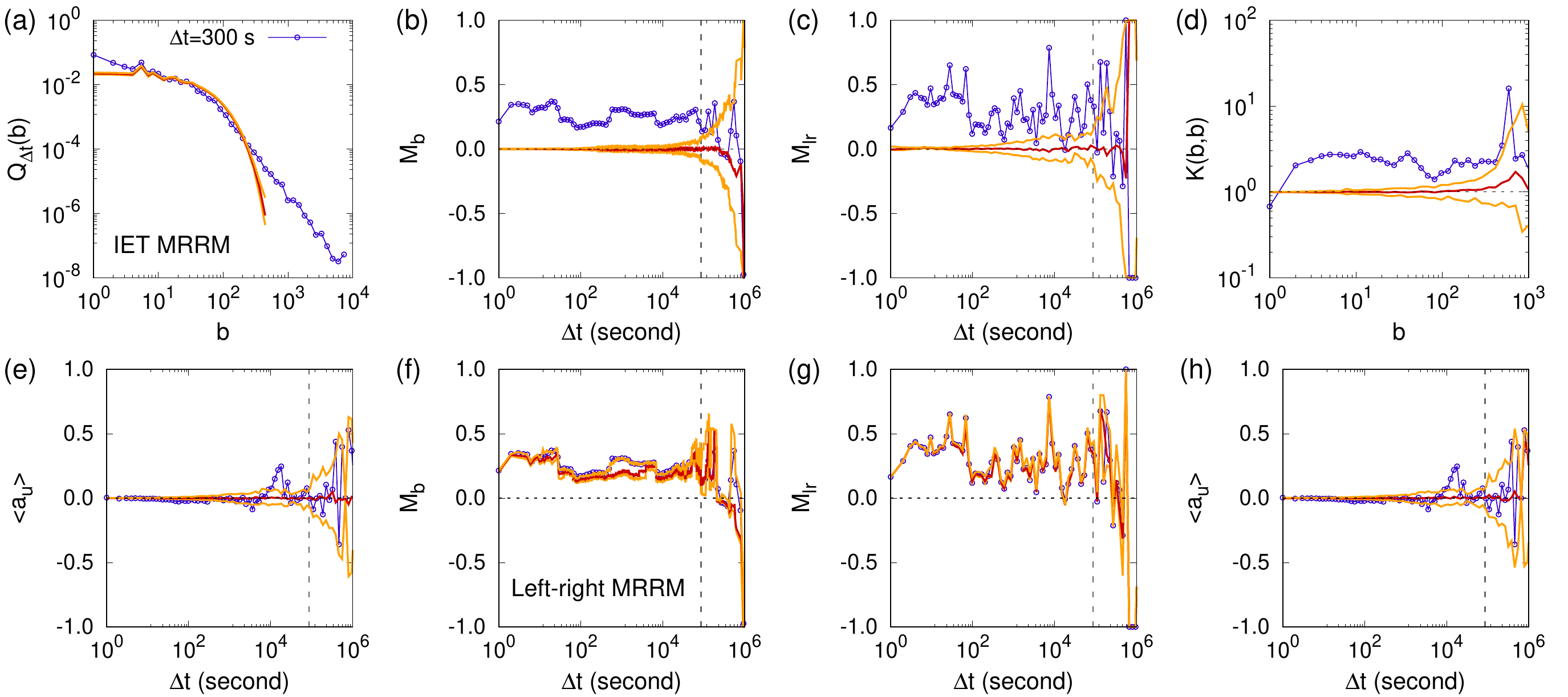}
    \caption{Results of two microcanonical randomized reference models (MRRMs) for the Wikipedia editor 1's edit sequence using $100$ randomized sequences for each MRRM: (a--e) IET MRRRM and (f--h) left-right MRRM. In all panels, the median is plotted by the red solid curve, while $95$th and $5$th percentiles by the orange solid curves. The original curves are also plotted by blue symbols for comparison. Vertical dashed lines in some panels denote 1 day.}
    \label{fig:editor1_rrm}
\end{figure*}

Our framework of the burst-tree decomposition can be straightforwardly applied to any other event sequences. We have analyzed edit sequences of other active editors in the English Wikipedia, tweet sequences of active Twitter users, heartbeat time series of healthy subjects, and the earthquake sequence in the JUNEC. Among them, the results for the Twitter user 1, for the heartbeat subject 1, and for the earthquake sequence are summarized in Fig.~\ref{fig:other_data}, while those for other Wikipedia editors, other Twitter users, and other heartbeat subjects are presented in Figs.~S2--S4 of Supplemental Material~\footnotemark[4]. For most event sequences analyzed, we find heavy-tailed burst size distributions $Q_{\Delta t}(b)$ for several values of $\Delta t$, positive $M_b$ and $M_{lr}$ for a wide range of $\Delta t$, merging kernels $K(b_v,b_w)$ with overall increasing $K(b,b)$, and negligible values of $\langle a_u\rangle$ for a wide range of $\Delta t$.

Interestingly, as shown in Fig.~\ref{fig:other_data}, we commonly observe nontrivial structure of burst trees, such as heavy-tailed burst size distributions, irrespective of the functional form of IET distributions $P(\tau)$. This clearly shows that the IET distributions and the burst-tree structures are not only separable---as is evident from the equivalence relation of $\mathcal{E} \widehat{=} (P(\tau),\mathcal{G})$---but there does not seem to be a strong connection between the IET distributions and the higher-order structures.

Finally, we remark that there could be alternative mechanisms behind the power-law distributions of burst sizes. Since the merging process with increasing $\Delta t$ is similar to the percolation process with increasing connectivity between elements of the system, one can hypothesize that the power-law burst size distribution could correspond to the power-law distribution of connected component sizes at the percolation transition point. However, this analogy may not be plausible because the power-law burst size distribution is observed for a wide range of $\Delta t$ in our empirical analysis, while the power-law distribution of connected component sizes appears only at the percolation transition point in the conventional percolation problem~\cite{Stauffer1994Introduction, Newman2010Networks}. By measuring the fraction of the largest burst size and the susceptibility as functions of $\Delta t$ for the editor 1, we find that the percolation transition occurs around at $\Delta t\approx 1$ day (see Sec.~III and Fig.~S1(c) in Supplemental Material~\footnotemark[4]), supporting our argument against the analogy to the percolation problem.

\subsection{Randomized reference models}\label{subsec:random}

Next, we demonstrate that our burst-tree decomposition method can be useful to systematically characterize temporal correlations or features in event sequences. For this, we adopt the methodology of microcanonical randomized reference models (MRRMs) for event sequences. The MRRMs have been extensively applied to characterize various features in temporal networks, see Ref.~\cite{Gauvin2018Randomized} and references therein. These MRRMs can be defined with the set of features they retain, and here we denote by $\mrrm{X}$ the MRRM which exactly retains the feature $X$, while maximally randomizing everything else. The MRRMs can be ordered according to the amount of information they preserve, such that the simpler or less informative the features are, the more of the original data is shuffled or randomized. In the case with event sequences, only very simple features discarding higher-order correlations such as the IET distribution have been considered in the literature~\cite{Gauvin2018Randomized}. To investigate the effects of keeping higher-order structures, compared to keeping only the simple ones, we will study the higher-order MRRMs based on the burst-tree structure.

The simplest MRRM used here is the one that only keeps the number of events, $\mrrm{n+1}$. This MRRM randomizes the timings of events by assigning to each event a random timing drawn from a uniform distribution in the entire time period $[t_0,t_n]$. In the limit of large $n$ this results in a Poisson process with an event rate determined only by the number of events and the time period~\cite{Kivela2012Multiscale}. The next simplest MRRM, denoted by $\mrrm{n+1,P(\tau)}$, retains the number of events and the IET distribution, which we call the IET MRRM. By permuting IETs in the empirical IET sequence, the IET MRRM only keeps correlations between two consecutive events, while all the higher-order correlations considering more than two consecutive events are destroyed.

\begin{table}[!t]
    \caption{Features or temporal correlations conserved in various microcanonical randomized reference models (MRRMs). The ``original'' data trivially retains all features. For definitions of MRRMs and symbols, see the main text.}
    \label{table}
    \begin{tabular}{lcccccc}
        \hline \hline
        MRRM & $P(\tau)$ & $Q_{\Delta t}(b)$ & $M_b$ & $M_{lr}$ & $K(b,b)$ &  $\langle a_u\rangle$ \\
        \hline
        Original & \checkmark & \checkmark & \checkmark & \checkmark & \checkmark & \checkmark \\
        Left-right shuffled & \checkmark & \checkmark & & & \checkmark & \\
        IET shuffled & \checkmark & & & & & \\
        Timing shuffled & & & & & & \\
        \hline \hline
    \end{tabular}
\end{table}

There are several possibilities of devising MRRMs based on the burst-tree structure. Here we are interested in the question about whether the arrow of time is relevant to the structure of temporal correlations. To study this issue, we first define an unoriented ordinal burst tree $\hat{\mathcal{G}}=\{(u, [v, w])\}$, which is the same as the ordinal burst tree $\mathcal{G}$ defined in the previous section, except that the set $[v,w]$ does not carry any information on the left-right orientation of $v$ and $w$. Now we devise an MRRM, denoted by $\mrrm{P(\tau),\hat{\mathcal{G}}}$, which keeps all features other than the left-right orientation of bursts. This MRRM is implemented by randomly swapping the left and right children for each internal node $u$, enabling us to call it the left-right MRRM. It strictly conserves the features measured by $P(\tau)$, $Q_{\Delta t}(b)$, and $K(b,b)$, while it may destroy other features measured by $M_b$, $M_{lr}$, and $\langle a_u\rangle$. In the case with $M_{lr}$ in Eq.~\eqref{eq:memory_lr}, the randomization keeps the average of $b_vb_w$, while it may change $\mu_l$, $\mu_r$, $\sigma_l$, and $\sigma_r$ but only marginally. Thus, we do not expect $M_{lr}$ in the left-right MRRM to be drastically different from the original result. The features conserved in the above three MRRMs are summarized in Table~\ref{table}~\footnote{For the Wikipedia editor 1 and Twitter user 1, one can also introduce variants of the left-right MRRM by limiting the shuffling only to internal nodes with IETs larger (or smaller) than a fixed timescale of $1$ day, aiming to destroy the left-right structure for the timescale longer (or shorter) than $1$ day. The results of these variants are included in Figs.~S5 and~S6 of Supplemental Material~[43].}.

Using the empirical event sequences, we test if the features that are not necessarily destroyed by the randomization still remain after the randomization. For each MRRM, we generate $100$ randomized event sequences to measure various quantities for detecting the corresponding features for each randomized event sequence, from which we obtain the curves of median, $95$th and $5$th percentiles to compare them to the original curves. As for the merging kernel, we compare the results of its diagonal cross-section $K(b,b)$ instead of $K(b_v,b_w)$ for effective comparison.

As an example, we find for the editor 1 that the shuffling of event timings destroys all temporal correlations, while the shuffling of IETs does the same apart from the IET distribution as expected, as shown in Fig.~\ref{fig:editor1_rrm}(a--e). Interestingly, the shuffling of the left-right children turns out to barely destroy the feature measured by $M_b$, while $M_{lr}$ is almost identical to the original result, as depicted in Fig.~\ref{fig:editor1_rrm}(f,~g). It implies that the correlations between two consecutive burst sizes (measured by $M_b$) might be dominated by those between sibling bursts (measured by $M_{lr}$). Finally, the time asymmetry measured by $\langle a_u\rangle$ is destroyed by the randomization, see Fig.~\ref{fig:editor1_rrm}(h). The complete results of the MRRM for the editor 1 are shown in Fig.~S5 of Supplemental Material~\footnotemark[4]. We also find the similar results for other datasets, i.e., the Twitter user 1, the heartbeat subject 1, and the JUNEC. For the complete results of MRRMs for the Twitter user 1, the heartbeat subject 1, and the JUNEC, see Figs.~S6--S8 of Supplemental Material~\footnotemark[4], respectively.

\begin{figure*}[!t]
    \includegraphics[width=0.6\textwidth]{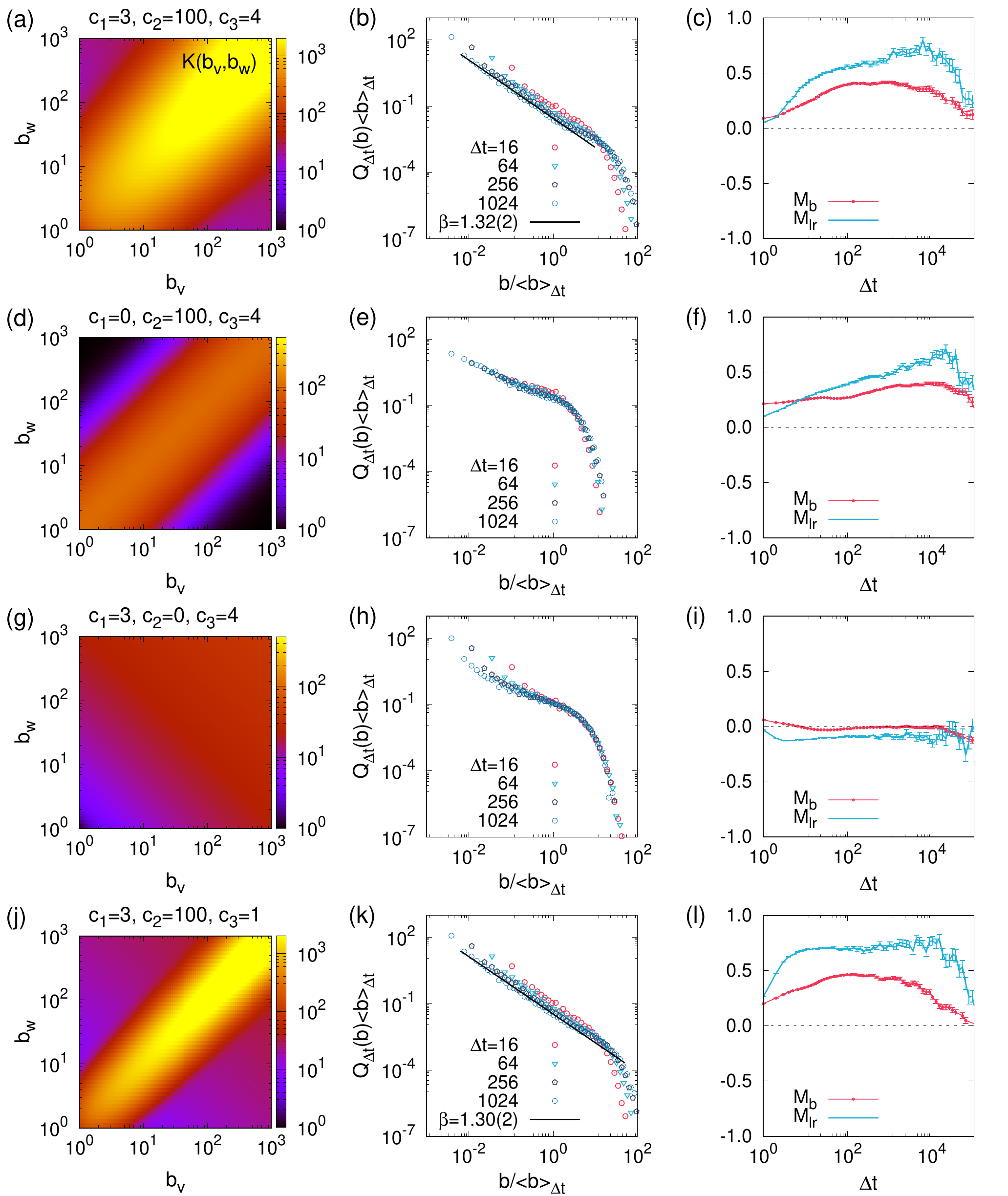}
    \caption{Simulation results of the kernel-based model using the model kernel in Eq.~\eqref{eq:model_kernel} with various sets of parameter values, as depicted in the left panels. For each case, $100$ event sequence of $n=10^5$ are generated, also using the IET distribution in Eq.~\eqref{eq:model_Ptau} with $\alpha=1.8$. By aggregating the detected burst sizes in $100$ event sequences, we obtain the burst size distributions $Q_{\Delta t}(b)$ for several values of $\Delta t$, rescaled by the average burst size $\langle b\rangle_{\Delta t}$ (center panels). The curves of memory coefficients $M_b$ and $M_{lr}$ are averaged over $100$ event sequences (right panels), where the error bars denotes the standard errors.}
    \label{fig:model}
\end{figure*}

The conclusion of the above MRRM study is that the features of the burst tree we observed for the data cannot be explained by randomness, but there is much more structure in the time series than just the IET distributions. Further, the temporal order of the bursts does not have a major effect on the observed burst-tree structure, apart from the asymmetries $\{a_u\}$ that we destroy in the left-right MRRM. Based on this conclusion, one can devise a simple model, mainly exploiting the merging kernel, for generating the event sequence with the empirically observed higher-order temporal correlations, as discussed in the next section.

\section{Kernel-based modeling}\label{sec:model}

Based on the empirical findings for the merging kernel, we can devise a simple model to reproduce the temporal correlations observed in the empirical event sequences. To generate an event sequence consisting of $n+1$ events, we need an interevent time (IET) distribution $P(\tau)$ for drawing $n$ IETs, and an ordinal burst tree $\mathcal{G}=\{(u,v,w)\}$ with $n$ internal nodes for the higher-order correlations between those IETs. For constructing the ordinal burst tree with $n$ internal nodes, we begin with $n+1$ events or leaf nodes, i.e., $n+1$ bursts of size $1$. Then two bursts, say $b$ and $b'$, are randomly chosen with a probability proportional to a model kernel $K(b,b')$. These two bursts are merged and randomly set as the left and right children of the merged burst, i.e., their parent node. This parent node is indexed by $n$ as this node will be associated with the smallest IET in $P(\tau)$. The next merging leads to another parent node to be indexed by $n-1$, and so on. This binary merging is repeated until we end up with the giant burst of size $n+1$. Inspired by the empirical merging kernels, e.g., in Figs.~\ref{fig:diagram_editor1}(f) and~\ref{fig:other_data}(d,~n), we adopt the following model kernel:
\begin{eqnarray}
    K(b,b') &=& \left[1+ c_1 (\ln b + \ln b')\right] \nonumber \\
    && \times \left[ 1+ c_2 e^{-(\ln b - \ln b')^2/c_3} \right]
    \label{eq:model_kernel}
\end{eqnarray}
with positive parameters $c_1$, $c_2$, and $c_3$. The first parenthesis of the right hand side in Eq.~\eqref{eq:model_kernel} describes an increasing behavior of the diagonal cross-section along the line of $b=b'$ for the preferential mixing of bursts. The second parenthesis is to implement the symmetrically decaying behavior with respect to the diagonal axis, which is for the assortative mixing of bursts. For example, see the heatmap of this model kernel for $c_1=3$, $c_2=100$, and $c_3=4$ in Fig.~\ref{fig:model}(a). 

Once the ordinal burst tree is ready, we draw $n$ random numbers from an IET distribution $P(\tau)$ to get a set of $n$ IETs, $\{\tau_i\}$. Precisely, we use a power-law IET distribution with an exponent $\alpha>1$:
\begin{equation}
    P(\tau)=(\alpha-1)\tau^{-\alpha}\ \textrm{for}\ \tau\geq \tau_{\rm min}=1. 
    \label{eq:model_Ptau}
\end{equation}
After assigning the IETs in $\{\tau_i\}$ to the internal nodes in $\mathcal{G}$, e.g., the largest IET to the root node, the event sequence of $n+1$ events is obtained by setting $t_0=0$ and then by calculating the event timings as $t_i=t_{i-1}+\hat\tau_{u(i)}$, where $u(i)$ denotes the $i$th visited internal node when traversing the ordinal burst tree in the inorder. This event sequence is analyzed to find that our simple kernel-based model successfully generates the event sequence showing the heavy-tailed burst size distributions for several values of $\Delta t$ as well as positive $M_b$ and $M_{lr}$ for a wide range of $\Delta t$, as shown in Fig.~\ref{fig:model}(b,~c). 

Using our kernel-based model, we can test if both preferential and assortative mixing structures are necessary for the power-law burst size distributions. The case with only assortative mixing can be studied by setting $c_1=0$, leading to the constant $K(b,b)$, as depicted in Fig.~\ref{fig:model}(d). This leads to thinner tails of $Q_{\Delta t}(b)$ than those for the case with $c_1>0$. Yet the positive $M_b$ and $M_{lr}$ are observed, implying that the assortative mixing is not sufficient to generate the power-law burst size distributions. Next, we consider the case only with the preferential mixing by setting $c_2=0$. Then the diagonal part of $K(b,b')$ is no longer higher than the off-diagonal part, as depicted in Fig.~\ref{fig:model}(g). It turns out that tails of $Q_{\Delta t}(b)$ are thinner than those for the case with $c_2>0$. Further, the values of $M_b$ and $M_{lr}$ are almost zero or even slightly negative for a wide range of $\Delta t$, because big and small bursts can be merged with each other more easily. Therefore, we conclude that both preferential and assortative mixing structures are necessary for obtaining the power-law $Q_{\Delta t}(b)$ and positive $M_b$ and $M_{lr}$ simultaneously. 

Finally, we test the effect of $c_3$ on the results: The smaller $c_3$ leads to the steeper decay of $K(b_v,b_w)$ along the direction perpendicular to the diagonal axis, as shown in Fig.~\ref{fig:model}(j). As the smaller $c_3$ would enhance the possibility of merging bursts of similar sizes, one can expect $M_{lr}$ to have the larger values than for the case with the larger $c_3$, which is indeed the case as shown in Fig.~\ref{fig:model}(l). We also find no considerable differences in $M_v$ as well as in $Q_{\Delta t}(b)$. In particular, the shapes of $Q_{\Delta t}(b)$ are quite similar to the case with the larger $c_3$, probably because the heavy tails of $Q_{\Delta t}(b)$ are largely affected by the characteristics of the diagonal cross-section $K(b,b)$.

We also have tested other functional forms of the model kernel to draw the qualitatively same conclusions, see Sec.~V and Figs.~S9 and~S10 of Supplemental Material~\footnotemark[4]. In addition, we remark that the asymmetric behavior between sibling bursts can be easily implemented in our model, e.g., by assigning a bigger (smaller) burst among chosen $b$ and $b'$ to the right (left) child with a probability $p$ ($q=1-p$). Then the case with $p=q=1/2$ reduces to our model, while the asymmetry can be implemented when $p,q\neq 1/2$.

\section{Conclusion}\label{sec:concl}

The comprehensive characterization of temporal correlations observed in various natural and social processes is crucial to the understanding of the underlying mechanisms behind such temporal processes. Non-Poissonian or bursty temporal patterns in empirical event sequences have been studied mostly in terms of heterogeneous interevent times (IETs), while the higher-order correlations between IETs are far from being fully understood due to the lack of the proper characterization method. In this paper we have proposed the burst-tree decomposition method that decomposes a given event sequence into the IET distribution and the ordinal burst tree, hence without loss of information on the temporal correlations. This implies that the ordinal burst tree, together with an IET distribution, can exactly reproduce the original event sequence. Using our burst-tree decomposition method one can systematically study the hierarchical structure of temporal correlations: In particular, the preferential and assortative mixing structure of bursts is empirically validated by measuring the novel memory coefficients between consecutive bursts and between sibling bursts as well as the merging kernel. In addition, the burst-tree decomposition turns out to be useful for the systematic investigation of temporal correlations in the framework of randomized reference models~\cite{Gauvin2018Randomized}. Finally, based on the empirically estimated merging kernels, we devise a kernel-based model to successfully generate event sequences showing the higher-order temporal correlations observed in the empirical datasets. 

We remark that once the ordinal burst tree is derived or given, it can be associated with any other set of IETs, irrespective of the functional form of the IET distribution. This implies that apparently very different event sequences might have the similar temporal correlation structure when their burst trees look similar to each other. We have observed this type of phenomenon in the empirical event sequences that show heavy-tailed burst size distributions despite of very different IET distributions. In addition, mapping the structure of temporal correlations onto a tree enables to propose other novel quantities for measuring various higher-order correlations as a tree structure is more intuitive and better visualized than the time series itself. Finally, we have considered only a binary tree in our work, while for more realistic decomposition of the event sequences in various datasets more complex trees than a binary tree can be used in a future.

\begin{acknowledgments}
    The authors thank Woo-Sik Son for providing us with the preprocessed dataset of the English Wikipedia.
    H.-H.J. and T.H. acknowledge financial support by Basic Science Research Program through the National Research Foundation of Korea (NRF) grant funded by the Ministry of Education (NRF-2018R1D1A1A09081919).
\end{acknowledgments}

\bibliographystyle{apsrev4-1}
%
    
\end{document}